\documentclass[12pt]{article}
\title{
Dynamical equations for mesons and baryons in large $N_c$ QCD
} \author{Yu.A.Simonov\\
State Research Center\\Institute of Theoretical and
Experimental Physics, \\
Moscow, Russia}
 \date{} \newcommand{\be}{\begin{equation}}
\newcommand{\ee}{\end{equation}}

\def\fun#1#2{\lower3.6pt\vbox{\baselineskip0pt\lineskip.9pt
\ialign{$\mathsurround=0pt#1\hfil
##\hfil$\crcr#2\crcr\sim\crcr}}}

\begin{document}
\maketitle

\begin{abstract}

 New equations are derived for meson and baryon Green's functions in
 a systematic method using the QCD Lagrangian and large $N_c$
 approximation as a starting point. The equations  are shown
 to contain both  confinement dynamics  and chiral symmetry
 breaking. Linear Regge trajectories are
 demonstrated for the solutions.
 \end{abstract}

\noindent{\large\bf 1.  Introduction}\\

\noindent
 The QCD dynamics of $q\bar q$ and $3q$ systems is governed
by two basic phenomena: confinement and chiral symmetry breaking
(CSB), which should be treated in a fully relativistically covariant
 way.  Confinement is usually introduced for static quarks via the
 area law of the Wilson loop [1] or equivalently through the field
 correlators in the Field Correlator Method (FCM) [2,3].

 For spinless quarks, or neglecting spin--dependent mass corrections,
 one can envisage a self--consistent method which treats confinement
 as the area law also for  light quarks in a relativistically
 covariant way. Such method was introduced originally in [4] for
 mesons, in [5] for baryons,  and in [6] for heavy--light mesons, and
 later on in  [7]  an improvement of the method was done taking into
 account dynamical degrees of freedom of the QCD string, which
 naturally appears due to the area law.

 As a result Regge trajectories have been found in [7] with the
 correct string slope $(2\pi \sigma)^{-1}$.

 Spin corrections have been considered in [8] for heavy mesons and in
 [6] for heavy--light ones, while baryon Regge trajectories have been
 found in [5], for review see [9].

 In all  cases the basic formalism was FCM and the Feynman--Schwinger
 (or world-line) path integral representation [3,10,11] which
 is well suited for relativistic quarks when spin is considered as a
 perturbation.

 The main difficulty which was always present in the method, was this
 perturbative treatment of spin degrees of freedom (which is
 incorrect, e.g., for pion) and absence of spontaneous CSB effects  in
 general [12].

 Recently a new type of formalism was suggested to treat
 simultaneously confinement and CSB and a new nonlinear equation was
 derived for a light quark in the field of heavy antiquark [13].

 This equation derived directly from QCD Lagrangian was found to
 produce linear confinement and CSB for the light quark [13,14] and
 the explicit form of the effective quark mass operator $M(x,y)$ was
 defined obeying both these properties.

 Since the method of [13] is quite general and allows to treat also
 multiquark systems, we apply it here to the $q\bar q$ and $3q$
 systems, and find dynamical equations for them, which contain
 confinement and CSB. To make our equations treatable, we
 systematically exploit the large $N_c$ limit, and mostly confine
 ourselves to the simplest field correlators -- the so--called
 Gaussian approximation;  it was shown in [13] that the sum
 over all correlators does not change the qualitative results,
 however the kernel of equations becomes much more  complicated.

 The paper is organized as follows.

 In Section 2 the general effective quark Lagrangian from the
 standard QCD  Lagrangian is obtained by integrating out gluonic
 degrees of freedom, and the nonlinear equation for the single quark
 propagator $S$ (attached to the string in a gauge--invariant way) is
 derived.

 Section 3 is devoted to the $q\bar q$ Green's function, which can be
 expressed as an integral of product of $S$ for quark and antiquark.
 Resulting equation is studied both in the differential form and in
 the integral form where again the Feynman--Schwinger representation
 leads to the effective Hamiltonian for the $q\bar q$ system,
 discussed in Appendix.

 A similar procedure is accomplished for the baryon Green's function
 in Section 4. Inclusion of chiral degrees of  freedom is given in
 Section 5, where an effective chiral Lagrangian is
 derived for bosons. Discussion of results and comparison to other
 methods is done in Conclusion.\\

\noindent{\large\bf 2.
 Effective quark Lagrangian}\\

\noindent
 As was discussed in the previous section, one can obtain effective
 quark Lagrangian by averaging over background gluonic fields. We
 shall repeat this procedure  following [13] now paying special
 attention to the dependence on the  contour in the  definition of
 contour gauge, and  introducing the operation of averaging over
 contour manifold.
  The QCD partition function for
 quarks and gluons can be written as
 \be
 Z=\int DAD\psi D\psi^+ {\rm exp} [L_0+L_1+L_{{\rm int}}]
 \ee
  where we are using Euclidean metric and define
  \be
  L_0=-\frac14\int d^4x(F^a_{\mu\nu})^2,
  \ee
  \be
  L_1=-i\int~^f\psi^+(x)(\hat \partial+m_f)~^f\psi(x)d^4x,
  \ee
  \be
  L_{\rm int}=\int~^f\psi^+(x) g\hat A(x)~^f\psi(x)d^4x.
  \ee

  Here and in what follows $~^f\psi_{a\alpha}$ denotes quark operator
  with flavour $f$, color $a$ and bispinor index $\alpha$.

  To express $A_\mu(x) $ through $F_{\mu\nu}$ one can use the
  generalized Fock--Schwinger gauge [15] with the contour $C(x)$ from
  the point $x$ to $x_0$, which can also lie at infinity,
  \be
  A_\mu(x)=\int_c F_{\lambda\beta} (z)
  \frac{\partial z_\beta (s,x)}{\partial x_\mu}
  \frac{\partial z_\lambda}{\partial s}
  ds.
  \ee
   Now one can integrate out gluonic field $A_\mu(x)$, and
  introduce an arbitrary integration over the set of contours $C(x)$
  with the weight $D_\kappa(C)$, since $Z$ is gauge invariant and
  does not depend on contours  $C(x)$. One obtains
  \be
  Z=\int D\kappa
 (C)D\psi D\psi^+{\rm  exp} \{L_1+L_{\rm eff}\}
 \ee
  where the
  effective quark Lagrangian $L_{\rm eff}$ is defined as
  \be
   {\rm
  exp} L_{\rm eff}=\langle {\rm exp} \int~^f\psi^+\hat A~^f\psi
  d^4x\rangle_A.
   \ee
    Using cluster expansion $L_{\rm eff}$ can be written
    as an infinite sum containing averages $\langle (\hat
   A)^k\rangle_A$.  At this point one can exploit the Gaussian
  approximation, neglecting all correlators $\langle (\hat
   A)^k\rangle$ for degrees  higher than $k=2$.  Numerical accuracy
   of this approximation was discussed in [9,16] and tested in [17].
   One expects that for static quarks corrections to Gaussian
   approximation amount to  less than 10\%.

   The resulting effective Lagrangian is quartic in $\psi$,
   \be
   L^{(4)}_{\rm eff}=\frac{1}{2N_c} \int d^4x
   d^4y^f\psi^+_{a\alpha}(x)~^f\psi_{b\beta}(x)~^g\psi^+_{b\gamma}(y)
   ~^g\psi_{a\delta}(y)J_{\alpha\beta;\gamma\delta}(x,y)+O(\psi^6),
   \ee
   \be
   J_{\alpha\beta,\gamma\delta}(x,y)=(\gamma_{\mu})_{\alpha\beta}
   (\gamma_\nu)_{\gamma\delta} J_{\mu\nu}(x,y)
   \ee
   and $J_{\mu\nu}$ is expressed as
   \be
   J_{\mu\nu} (x,y)=g^2\int^x_C\frac{\partial
   u_\omega}{\partial x_\mu} du_\varepsilon \int^y_C \frac{\partial
   u_{\omega'}}{\partial  y_\nu}
   du_{\varepsilon'}\frac{tr}{N_c}\langle
   F_{\varepsilon\omega}(u)
   F_{\varepsilon'\omega'}(v)\rangle.
   \ee
$L_{\rm eff}$ (8) is written in the contour gauge [15].

   It can be identically rewritten in the
   gauge--invariant form if one substitutes parallel
   transporters $\Phi(x,x_0),\Phi(y, x_0)$ (identically
   equal to unity in this gauge) into (8) and (10),
   multiplying each $\psi(x)$ and $\psi(y)$ respectively
   and in (10) replacing $F(u)$ by
   $\Phi(x,u)F(u)\Phi(u,x_0)$ and similarly for $F(v)$.

   After that $L_{\rm eff}$ becomes gauge--invariant, but in
   general contour--dependent, if one keeps only the
   quartic term (8), and neglects all higher terms. A
   similar problem occurs in the cluster expansion of
   Wilson loop, when one keeps only lowest correlators,
   leading to the (erroneous) surface dependence of the
   result.

   Situation here is the same as with a sum of QCD
   perturbation series, which depends on the
   normalization mass $\mu$ for any finite number of
   terms in the series. This unphysical dependence is
   usually treated by fixing $\mu$ at some physically
   reasonable value $\mu_0$ (of the order of the inverse
   size of the system.

   The integration over contours $D\kappa(C)$ in (6)
   resolves this difficulty in a similar way.
      Namely, the partition function $Z$ formally does not
   depend on contours (since it is integrated over a set
   of contours) but depends on the  weight
   $D\kappa(C)$ and we choose this weight in such a way,
   that the contours would generate the string of
   minimal length between $q$ and $\bar q$. Thus the
   physical choice of the contour corresponds to the minimization of
   the meson (baryon) mass over the class of strings,
   in the same way as the choice of $\mu=\mu_0$
   corresponds to the minimization of the dropped higher
   perturbative terms.

   As a practical outcome, we shall keep the integral
   $D\kappa(C)$ till the end
   and finally use it to minimize the string between the quarks.

   Till this point we have made only one approximation --neglected
   all field correlators except the Gaussian one. Now one must do
   another approximation --assume large $N_c$ expansion and keep the
   lowest term. As was shown in [13] this enables one to replace in
   (8) the colorless product $~^f\psi_b(x)~^g\psi_b^+(y)=
   tr (~^f\psi(x)\Phi(x, x_0)\Phi(x_0,y)~^g\psi^+(y))
   $
   by  the quark Green's function
   \be
   ~^f\psi_{b\beta}(x)~^g\psi^+_{b\gamma}(y)\to
   \delta_{fg}N_cS_{\beta\gamma}(x,y)
   \ee
   and $L^{(4)}_{\rm eff}$ assumes the form
   \be
   L^{(4)}_{\rm eff}=-i\int d^4xd^4y~^f\psi^+_{a\alpha}(x)
   ~^fM_{\alpha\delta}(x,y) ~^f\psi_{a\delta}(y)
   \ee
   where the quark mass operator is
   \be
   ~^fM_{\alpha\delta}(x,y)=-J_{\mu\nu}(x,y) (\gamma_{\mu}~^fS(x,y)
   \gamma_\nu)_{\alpha\delta}.
   \ee
   From (6) it is evident that $~^fS$ satisfies to  equation
   \be
   (-i\hat \partial_x-im_f)~^fS(x,y)-i\int~^fM(x,z) d^4
   z~^fS(z,y)=\delta^{(4)}(x-y).
   \ee
   Equations (13), (14) have been first derived in [13]. From (6) and
   (12) one can realize that at large $N_c$ the $q\bar q$ and $3q$
   dynamics is expressed through the quark mass operator (13), which
   should contain both confinement and CSB.

   Indeed, analysis done in [13,14] reveals that confinement is
   present in the long--distance form of $M(x,y)$, when both
   distances $|\mbox{\boldmath ${\rm  x}$}|, |\mbox{\boldmath ${\rm
   y}$}|$ of light quark from heavy antiquark (placed at
   $\mbox{\boldmath $ {\rm x}$} =0$) are large.

   We shall  do now several simplifying assumptions, to clarify the
   structure of $M(x,y)$. First of all we take the class of contours
   $C$ going from any point $x=(x_4, \mbox{\boldmath $ {\rm x}$})$ to
   the point $(x_4,0)$ and then to $(-\infty)$ along the $x_4$ axis.
   For this class the corresponding gauge was studied in [18].
   Secondly, we take the dominant part of $J_{\mu\nu}$ in (13),
   namely $J_{44}$, which is proportional to the correlator of
   color--electric fields, yielding linear confining interaction,
   and neglect other components $J_{ik}, J_{i4}, J_{4i}, i=1,2,3$,
   containing magnetic fields and yielding momentum dependent
   corrections. (It is easy to take into account these contributions
   in a more detailed analysis).

   The correlator $\langle FF\rangle$ in (10) can be expressed
   through the scalar correlator $D(x), $ defined as [2],
   \be
   \frac{trg^2}{N_c}\langle F_{\alpha\beta}(u) \Phi(u,v)
   F_{\gamma\delta} (v)\Phi(v,u)\rangle =
   D(u-v)(\delta_{\alpha\gamma} \delta_{\beta\delta}
   -\delta_{\alpha\delta}\delta_{\beta\gamma})+O(D_1)
   \ee
   where the correlator $D_1$ not contributing to confinement is
   neglected. As a result one has for $M$ [19]
   \be
   ~^fM_{C_{x_4}}(x,y)=
   ~^fM^{(0)}I+~^fM^{(i)}\hat \sigma_i
   +~^fM^{(4)}\gamma_4+~^fM^{(i)}_\gamma \gamma_i.
   \ee
   Here we have  defined
\be
\hat \sigma_i=\left(
\begin{array}{ll}
\sigma_i&0\\
0&\sigma_i
\end{array}\right).
\ee
The dominant part of $M$, $~^fM^{(0)}$ is linearly growing at large
$|\mbox{\boldmath ${\rm x}$}|, |\mbox{\boldmath ${\rm y}$}|$ and in
the most simple case of Gaussian form of $D(x)$, can be written as
\be
 ~^fM^{(0)}(x,y)=\frac{1}{2T_g\sqrt{\pi}}
e^{-\frac{(x_4-y_4)^2}{4T_g^2}}\sigma|\frac{\mbox{\boldmath ${\rm x}$}
+\mbox{\boldmath ${\rm y}$}}{2}|\tilde \delta ^{(3)}
(\mbox{\boldmath
${\rm x}$}-\mbox{\boldmath${\rm y}$})
 \ee
  where   $T_g$ is the gluon corretion length, and
   $\tilde \delta$ is a
smeared $\delta$--function, which can be represented as [19] \be
\tilde \delta^{(3)}
(\mbox{\boldmath
${\rm x}$}-\mbox{\boldmath${\rm y}$})
\approx exp (-\frac{|
\mbox{\boldmath
${\rm x}$}-\mbox{\boldmath${\rm y}$}
|^2}{b^2})(\frac{1}{b\sqrt{\pi}})^3,~~b\sim 2T_g.
\ee
Here $T_g$ is the gluon correlation length, which enters $D(u)$ as
$D(u)= D(0) {\rm exp}(-\frac{u^2}{4T_g^2})$. We are now in the
position to derive $q\bar q$,$ 3q$ Green's function, which will be
done in the next section. \\

\noindent{\large\bf 3.
Equations for $q\bar q$  Green's function}\\

\noindent
 We start with gauge--invariant definitions of initial
 ($\Psi_{\rm in}$) and final ($\Psi_{\rm fin})~~ q\bar q$ states,
 $$ \Psi_{\rm in}(x,\bar x)\equiv ~^{f_1}\psi^+(x)\Gamma_{f_1 \bar
 f_1}^{\rm in}\Phi(x,\bar x)~^{\bar f_1}\psi(\bar x); ~~ $$ \be
 \Psi_{\rm fin}(y, \bar y)= ~^{f_2}\psi^+(y)\Gamma_{f_2
 \bar f_2}^{\rm fin}\Phi(y,\bar y)~^{\bar f_2}\psi(\bar y).
 \ee
 The $q\bar q$ Green's function is expressed as
$$
 G(x\bar x|y\bar y)=\langle \Psi_{\rm in}(x,\bar x)
 \Psi^+_{\rm fin}(y, \bar y)\rangle= $$ \be = \frac1N\int
 D\kappa(C)D\psi D\psi^+ {\rm exp} \{
 L+L_{\rm eff}\}\Psi_{\rm in}\Psi^+_{\rm fin}. \ee

    One can do integrals over $D\psi D\psi^+$ and assuming as before
    the large $N_c$ limit, one can neglect the determinant term,
    yielding
 $$
 G(x\bar x|y\bar y)=
 \frac1N\int D\kappa(C)\{ {\rm tr}[
 S_{\bar C}^{\bar f_1}(\bar x, \bar y)
 \Gamma^{+{\rm fin}}_{f_1\bar f_2}
 S_{\bar C}^{ f_2}(y,x)
 \Gamma^{\rm in}_{f_2\bar f_1}]-
 $$
 \be
 -tr (S_C^{f_1}(x,\bar x)\Gamma^{\rm in})
 {\rm tr} (S_C^{\bar f}(y,\bar y)\Gamma^{+{\rm fin}})\}
 \ee
 where we have omitted for simplicity parallel transporters and used
 notation
 \be
 S_C^{f}(x,y)=
 (-i\hat \partial -im_f-iM^f(x',y'))^{-1}_{x,y}.
 \ee
 At this point we assume that minimization over contours $D\kappa(C)$
 in (22) selects one specific contour $\tilde C(u)$ for each $u$ (and
 we find this $\tilde C(u)$ in what follows), and choose for
 simplicity nonsinglet flavour channel.
  Then the second term in the curly brackets in (22) disappears and
  we have equation for $G$
 $$
  (-i\hat \partial_x-im_1-iM_{C_1}(x,y))(-i\hat \partial_{\bar x}
  -im_2-iM_{C_2}(\bar x, \bar y)) G(y,\bar y|z,\bar z)=
  $$
  \be
   \delta^{(4)}(x-z)\delta^{(4)}(\bar x-\bar z)
  \ee
  where we have  omitted integrals over $d^4y d^4\bar y$ and unit
  operators $\hat 1=\delta^{(4)}(x-y), \hat{\bar 1}=\delta^{(4)}(\bar
  x-\bar y)$ as factors of
  $(-i\hat \partial_x-im_1)$ and
  $(-i\hat \partial_{\bar x}-im_2)$, respectively.
  Next steps in treating (24) are standard for the   Bethe--Salpeter
 formalism [20], the only difference is the contour dependence of
 mass operators $M_{C_i}$, which we specify as follows. One can
 choose the class of contours which pass from a point $u=(u_4,
\mbox{\boldmath
${\rm u}$})$ to $(u_4,
\mbox{\boldmath
${\rm x}$}_0)$  and then to $(-\infty,
\mbox{\boldmath
${\rm x}$}_0)$, i.e.
 parallel to the  c.m. motion of the $q\bar q$ system. Exact value of
 $
\mbox{\boldmath
 ${\rm x}$}_0$ will be found by minimization of the
 interaction kernel.

 Defining as in [20] the c.m. coordinate $X_{\mu}$ and
 relative coordinate $r_\mu$.
 \be
 X_\mu=\alpha x_\mu + (1-\alpha)\bar x_\mu, r_\mu=
 x_\mu-\bar x_\mu,
\ee
where $\alpha$ is an  arbitrary parameter and
corresponding  momenta
\be
P_{\mu}=\frac{1}{i}\frac{\partial}{\partial X_\mu}, ~~
p_\mu=\frac{1}{i}\frac{\partial}{\partial r_\mu}
\ee
one can also fix $\alpha$ to have correspondence with
nonrelativistic limit to be $\alpha={m_1}({m_1+m_2})^{-1}$.

Since $M_C$ is invariant with respect  to time shifts,
one can introduce the  total mass of the $q\bar q$
system $E$
 and relative energy $\varepsilon$ and write in the c.m. system an
 equation for the $q\bar q$ wave function
  $$
[\frac{m_1}{m_1+m_2}E-(m_1\beta_1+\mbox{\boldmath
  ${\rm p}$}\cdot
\mbox{\boldmath
  ${\rm \alpha}$}_1)+\varepsilon - \beta_1 U(
\mbox{\boldmath
  ${\rm x}$}-
\mbox{\boldmath
  ${\rm x}$}_0)]
  $$
  \be
  \times
[\frac{m_2}{m_1+m_2}E-(m_2\beta_2-
\mbox{\boldmath
${\rm  p}$} \cdot
\mbox{\boldmath
${\rm \alpha}$}_2)-\varepsilon - \beta_2 U(
\mbox{\boldmath
${\rm x}$}-
\mbox{\boldmath
${\rm x}$}_0)]\Psi=0
\ee
where $U(
\mbox{\boldmath
${\rm  x}$}-
\mbox{\boldmath
${\rm  x}$}_0)$ is the local limit $(T_g\to 0)$ of the mass
operator $M_C$. For vanishing angular momentum one has
\be
U(
\mbox{\boldmath
${\rm  x}$}-
\mbox{\boldmath
${\rm x}$}_0)=\sigma |
\mbox{\boldmath
${\rm  x}$}-
\mbox{\boldmath
${\rm x}$}_0|.
\ee

The further analysis and solution of (27) involves as in [21]
introduction of positive  (+) and  negative (-) energy projection
operators and corresponding 4 wave  functions
$\Psi_{\lambda\lambda'}, \lambda, \lambda'=\pm$, for which  a
 system of equations is written similarly to [21].

We shall not follow this route here, however, leaving it for
subsequent paper and instead write in the Appendix the Hamiltonian
approach to the same problem, which will allow us to calculate
spectrum in an approximate way, and discuss it in the Conclusions.

There is still another the 3d form of (27) which obtains in the same
way as in [21]. Indeed, writing the time Fourier trasform of the
Green's function in the form (the sign $\langle \rangle_C$ denotes
the integral $D\kappa(C)$)
\be
  G=\langle\frac{1}{(E-E_2-H_1)(E_2-H_2)}\rangle_C
\ee
 and integrating over $dE_2$ one arrives as in [21] to the 3d
Green's function
 \be
  G(E, \mbox{\boldmath
 ${\rm r}$},
\mbox{\boldmath
 ${\rm r}$}')=\langle\frac{1}{E-H_1-H_2}\rangle _C
\ee
 where
  $$
H_1=m_1\beta_1+
\mbox{\boldmath
${\rm p}$}\cdot
\mbox{\boldmath
${\rm \alpha}$}_1 +\beta_1 U(
\mbox{\boldmath
${\rm  x}$}-
\mbox{\boldmath
${\rm  x}$}_0),
$$
  \be
H_2=m_2\beta_2-
\mbox{\boldmath
$ {\rm p}$}\cdot
\mbox{\boldmath
${\rm  \alpha}$}_2 +\beta_2 U(\bar {
\mbox{\boldmath
${\rm x}$}}-
\mbox{\boldmath
${\rm  x}$}_0).
  \ee

   Minimizing interaction over $x_0$,
  one arrives
    at $x_0=\frac{
\mbox{\boldmath
${\rm x}$}+\bar{
\mbox{\boldmath
 ${\rm x}$}}}{2}$,
  and  at the equation for the wave
function.
  \be
   \{m_1\beta _1+m_2\beta_2+\mbox{\boldmath${\rm  p}$}(
\mbox{\boldmath
   ${\rm \alpha}$}_1-
\mbox{\boldmath
${\rm \alpha}$}_2) +\beta_1 U(\frac{
\mbox{\boldmath
${\rm  r}$}}{2})+\beta_2 U(-\frac{
\mbox{\boldmath
${\rm r}$}}{2})\} \Psi(\mbox{\boldmath
${\rm r}$}
)= E\Psi(
\mbox{\boldmath
${\rm  r}$}).
 \ee

    In the nonrelativistic limit this reduces to the usual
    nonrelativistic quark model,
    \be
    (\frac{
\mbox{\boldmath
    ${\rm p}$}^2}{2m}+\sigma r) \varphi_n=\varepsilon_n \varphi_n,
    ~\varepsilon_n =E_n-m_1-m_2, ~m=\frac{m_1m_2}{m_1+m_2}
    \ee \\

\noindent{\large\bf 4.
    Equations for the baryon Green's function}\\

\noindent
    Equations for the $3q$ system can be written in the same way as
    for the $q\bar q$ system.
    We again shall assume large $N_c$ limit in the sense, that
    $1/N_c$ corrections from $q\bar q$ pairs to the quark Green's
    function and the effective mass can be neglected, but shall write
    explicit expressions for $N_c=3$.

    The initial and final field operators are
    \be
    \Psi_{in}(x,y,z) = e_{abc}
    \Gamma^{\alpha\beta\gamma}\psi_{a\alpha}(x,C(x))
    \psi_{b\beta}(y,C(y))\psi_{c\gamma}(z, C(z))
    \ee
    with the notations: $a,b,c,$ are color indices, $\alpha,\beta,
    \gamma$ are Lorentz   bispinor indices and transported quark
    operators are
    \be
    \psi_{a\alpha}(x,C(x))=(\Phi_C(x,\bar x)\psi_{\alpha}(\bar
    x))_a
    \ee
    and the contour $C(x)$ in $\Phi_C$ can be  arbitrary , but it is
    convenient to choose  it in  the same class of contours that is
    used in $D\kappa (C)$ and  in the generalized Fock--Schwinger
    gauge [15]. $\Gamma^{\alpha \beta \gamma}$ is the  Lorentz spinor
    tensor securing proper baryon quantum numbers.  One can also
    choose other operators, but it does not influence the resulting
    equations.
    In (34) we have omitted flavour indices in $\Gamma$ and
    $\psi(x,C)$, to be easily restored in final expressions.

    Using now the effective Lagrangian (12) valid at large $N_c$, we
    obtain for the $3q$ Green's function.
    $$
    G^{(3q)}(x,y,z|x',y',z')=
    $$
    \be
   \frac1N\int D\kappa(C)D\psi
    D\psi^+\Psi_{\rm fin}(x',y',z')\Psi^+_{\rm in}(x,y,z)
    {\rm exp} (L_1+L_{\rm eff}).
     \ee
     Integrating out quark degrees of
    freedom and neglecting determinant at large $N_c$ one has \be
    G^{(3q)}=\int D\kappa(C) (e\Gamma) (e'\Gamma')\{S(x,x') S(y,y')
    S(z,z')+ {\rm perm}\} \ee where color and bispinor indices are
    suppressed for simplicity together with parallel transporters in
    initial and final states.

    One can also define unprojected (without $\Gamma,\Gamma')$ $3q$
    Green's function $G^{(3q)}_{\rm un}$ with 3 initial and 3 final
    bispinor indices instead of projected by $\Gamma, \Gamma'$
    quantum numbers of baryon.

    Assuming that minimization over contours $D\kappa(C)$ reduces to
    the single choice of the contours (the single string junction
    trajectory minimizing the mass of  baryon), one can write
    equation for $G_{\rm un}^{(3q)}$:
    $$
    (-i\hat\partial_x-im_1-i\hat M_1)
    (-i\hat\partial_y-im_2-i\hat M_2)
    (-i\hat\partial_z-im_3-i\hat M_3)
    G_{\rm un}^{(3q)}=
    $$
    \be
    \delta^{(4)}(x-x')
    \delta^{(4)}(y-y')
    \delta^{(4)}(z-z')
    \ee
    and e.g. $\hat M_1 G\equiv \int M(x,u) G(u,x') d^4 u$.
    One can simplify the form (37) for $G^{(3q)}$ taking into account
    that $M(x,x')$ actually does not depend on $\frac{x_4+x'_4}{2}$,
    and hence the interaction kernel of $G^{(3q)}$ does not depend on
    relative energies, as in [21]. Similarly to [20,21] one can
    introduce Fourier tranform of $G^{(3q)}$ in time components and
    take into account energy conservation $E=E_1+E_2+E_3$. One
    obtains
 $$
    G^{(3q)}(E,E_2, E_3)\simeq \int D\kappa(C) (e\Gamma) (e'\Gamma')
   $$
   \be
   \times
    \frac{1}{(E-E_2-E_3-H_1)(E_2-H_2)(E_3-H_3)}
    \ee
    where notations are used
    \be
    H_i=m_i\beta^{(i)}+
\mbox{\boldmath
    ${\rm  p}$}^{(i)}
\mbox{\boldmath
    ${\rm  \alpha}$}^{(i)}+
    \beta^{(i)}M(
\mbox{\boldmath
    ${\rm  r}$}^{(i)}-
\mbox{\boldmath
    ${\rm r} $}^{(0)})
    \ee
    and we have taken in $M(x,x')$ the limit of small $T_g$ and set
    of contours in $D\kappa(C)$ passing from the point $
\mbox{\boldmath
    ${\rm  r}$}^{(i)}$
    to some (arbitrary) point $
\mbox{\boldmath
    ${\rm r}$}^{(0)}$.

    As in [21] one can now integrate over $E_2, E_3$ to obtain
    finally
    \be
    G^{(3q)}(E,
\mbox{\boldmath
    ${\rm r}$}_i,
\mbox{\boldmath
    ${\rm r}$}'_i)\simeq \int D\kappa(C)
    (e\Gamma) (e'\Gamma') \frac{1}{(E-H_1-H_2-H_3)}.
    \ee
    From (41) one obtains equation for the $3q$ wave function similar
    to that of $q\bar q$ system,
        \be
    (H_1+H_2+H_3-E)\psi(
\mbox{\boldmath
    ${\rm r}$}_1,
\mbox{\boldmath
    ${\rm r}$}_2,
\mbox{\boldmath
    ${\rm r}$}_3)=0
    \ee
    where $\mbox{\boldmath ${\rm r}$}^{(0)}$ is to be taken at the
    Torricelli point.

            In the nonrelativistic approximation $m_i\gg
    \sqrt{\sigma}$ one has
    \be
    \sum^3_{i=1}\left [\frac{(
\mbox{\boldmath
    ${\rm  p}$}^{(i)})^2}{2m_i}+
    \sigma|
\mbox{\boldmath
    ${\rm r}$}^{(i)}-
\mbox{\boldmath
    ${\rm  r}$}^{(0)}|\right ] \Psi= \varepsilon
    \Psi~~ \varepsilon =E-\sum m_i
    \ee\\

\noindent{\large\bf 5.
    Inclusion of chiral degrees of freedom}\\

\noindent
    Analysis of the $q\bar q$ equations, e.g. (24) and (32), reveals
    that solutions do not contain Goldstone modes in the chiral limit
    ($m_1, m_2\to 0)$. The reason is that the equations have been
    obtained in the limit of large $N_c$, while chiral corrections
    appear in the subleading order, and the coupling constant of pion
    to the quark is $O(N_c^{-1/2} )$. Therefore chiral correction
    $O(1/N_c)$ to the  second term on the r.h.s. of (22) yields
    contribution of the same order $O(N_c)$ as the first term, since
    number of color traces in these terms are different. A similar
    situation occurs in the vacuum made of instantons surrounded by
    confining background considered in [22].  There the massive
    $\rho$--type pole in the pionic channel produced by the first
    term in (22) is exactly cancelled by the second term, while the
    Goldstone pole appears in the  chiral correction.

    To proceed one can bosonize the quartic Lagrangian (8) in the
    standard (but nonlocal) way [23]
    \be
    Z=\int D\kappa (C) D\psi d\psi^+ D\omega {\rm exp} \{
    L_1+L_\omega+L_{\psi \omega}\}
    \ee
    where bosonic colorless field $\omega \equiv
    \omega_{\alpha\beta}^{fg} (x,y)$ enters in
    $L_\omega,L_{\psi\omega}$ as
    \be
    L_\omega =-\frac{1}{2N_c} \int d^4 x d^4y \omega^{fg}_{\alpha
    \beta} (x,y) J_{\alpha \beta; \gamma \delta} (x,y)
    \omega^{fg}_{\gamma\delta} (x,y)
    \ee
    \be
    L_{\omega\psi} =-\frac{1}{2N_c} \int d^4 x d^4y
    [\omega^{fg}_{\alpha \beta} (x,y) J_{\alpha \beta; \gamma \delta}
    (x,y)~^f\psi^+_\gamma(x)~^g\psi_\delta(y)+{\rm h.c.}].
    \ee
   Integrating out quark degrees of freedom  one obtains the
   effective chiral Lagrangian, $Z\sim \int e^{L_{\rm ch}}D\omega
   D\kappa (C)$
   \be
   L_{\rm ch}=-L_\omega +{\rm Tr ln} (i\hat \partial +i\hat m+iM
   +\Delta) \ee
    where $M$ is the leading $O(N_c^0)$ term (13), while
   $\Delta$ is the chiral correction.

   Note that in the limit $T_g\to 0$, the quark mass operator $M$
   becomes   local $(\sim \delta^{(4)} (x-y))$ and therefore one can
   consider in $\Delta$ and $L_{\rm ch}$ the usual limit of local
   chiral field. Parametrizing it as in [24], one has
    \be
     \Delta_\pi
   =i(e^{i \mbox{\boldmath ${\rm  \pi}$} \mbox{\boldmath${\rm
   \tau}$}\gamma_5}-1) M.
   \ee
    Proceeding as in [22] one can expand
   $L_{\rm ch}(\pi)$in $\pi_i$ and obtain
    \be
    L_{\rm ch} (\pi) =\int
   \pi_a(k) N(k, k') \pi_a(k') dk dk' \ee where $N(0,0)=0$ yielding
   massless Nambu--Goldstone pions.  Similarly to [22] also here the
   massive pole cancels while the Nambu--Goldstone pole survives.
   For a detailed discussion the reader is referred to [22] and a
   subsequent publication.\\

\noindent{\large\bf 6.
   Conclusions}    \\

\noindent
   We have obtained above equations for mesons and baryons (27),
   (32) and (38), (42) respectively which contain all dynamics in the
   form of nonlocal mass Kernels $M(x,y)$. The latter are solutions
   of equations (13), (14) obtained earlier in [13] and  describing
   motion of one quark in the field of heavy antiquark. This
   reduction of $q\bar q$ and $3d$ problems to  a simpler problem of
   one quark in the field of static source is possible in the lowest
   order of $1/N_c$ expansion.

   The form of solution for $M(x,y)$ was found quasiclassically in
   [13,14]  and also numerically in [19]  in the thin string limit
   $T_g\to 0)$ and for nonrotating string $(L=0)$ reduces to a
   simple linear potential at large distances.

   In the same limit the obtained equations produce reasonable
   spectrum, see, e.g., equation (A.12) for the $q\bar q$ system,
   which yields results roughly coinciding with the WKB spectrum for
   spinless Salpeter equation in a case of zero current quark mass
   [25].

   Thus confining dynamics is correctly reproduced by the cited
   equations. At the same time the exploited form of  $M(x,y)$ is
   applicable at zero angular momentum and does not describe the
   rotating string, which can be seen in  the wrong Regge slope in
   (A.12). Chiral degrees of freedom which can be taken into account
   in the same formalism need additional step, which was shortly
   discussed in Section 5.  Here one should take into account the
   $4q$ form of the effective Lagrangian (8) and do a general
   (nonlocal) bosonisation procedure, which generates a new operator
   $\hat M(x,y)$, containing bosonic fields, in addition to the
   scalar--isoscalar component of $\hat M$, dominant at large $N_c$,
   which represents the scalar string (and hence chiral symmetry
   breaking).

   The paper has left aside many interesting questions,  in
   particular the exact form of equations for pion Green's function,
   the coexistence of  confining and chiral effects in the dynamics
   of mesons and baryons, the definition of the constituent quark
   mass and hadron magnetic moments etc.

   All these points are planned for future publications.

   The author is grateful to Yu.S. Kalashnikova,  A.V.Nefediev and
   V.I.Shevchenko for useful discussions.

   The financial support of RFFI through the grants 97-02-16404,
   97-02-17491 and 96-15-96740 is gratefully acknowledged.
   \newpage

\setcounter{equation}{0} \def\theequation{A.\arabic{equation}}

 \vspace{1cm}

\noindent
{\bf Appendix\\
\noindent
Effective Hamiltonian for the $q\bar q$ system}\\

\noindent
Our purpose here is to use the Feynman--Schwinger representation
(FSR) [10,11] to express the inverse operator in (23). To this end
we represent $S_C^f$ as
 \be S_C^f(x,y)= i (\hat \partial +m +M)^{-1}=
 i (-\hat \partial +m +M)
  (- \partial^2 +\delta +M^2)^{-1}
  \ee
  where $\delta=\hat \partial M-M\hat \partial$, and with the
  notation $\tilde M^2\equiv M^2 +\delta$  one has using FSR
$$
S_C^f(x,y)= i(-\hat \partial +m+M)\int^\infty_0 ds
e^{-s(-\partial^2+\tilde M^2)} =
$$
\be
 i (-\hat
\partial +m +M)\int^\infty_0 ds (Dz)_{xy} e^{-K-\int^s_0 d\tau\tilde
M^2(z\tau)}.
\ee
Proceeding now in the same way, as in [7] one obtains
$$
G(x,\bar x|y,\bar y) =\frac1N\int D\kappa(C) \Gamma^+
(-\hat \partial +m_{f_1}+M_{f_1})\Gamma
(-\hat \partial +m_{\bar f_1}+M_{\bar f_1})
$$
\be
\times
\int DR Dr D{\mu_1}
D{\mu_2} e^{-A}
\ee
where we have used notations
\be
R_\alpha=\frac{\mu_1z_\alpha+\mu_2\bar z_\alpha}{\mu_1+\mu_2},~~
r_\alpha =z_\alpha (t) -\bar z_\alpha (t), t\equiv z_4,
\ee
$$
2\mu_1 = \frac{dz_4}{d\tau}, 2\mu_2 =\frac{d\bar z_4}{d\bar
\tau}.
$$
The  Euclidean action $A$ can be written as
$$
A= \int^T_0 dt \left\{
 \frac{m_1^2}{2\mu_1(t)}+
 \frac{m_2^2}{2\mu_2(t)}+
 \frac{\mu_+(t)}{2} \dot{R^2}+
 \frac{\tilde \mu (t)\dot {
\mbox{\boldmath
 $ {\rm r}$}}^2}{2} +  \right.
 $$
 \be
 \left.
 +\frac{\tilde M_c^2 (z(t)-x_0)}{2\mu_1}+
 \frac{\tilde{\bar M}_{\bar c}^2 (\bar z(t)-x_0)}{2\mu_2}
 \right \};
 \ee
 with the notations $\tilde \mu=\frac{\mu_1\mu_2}{\mu_1+\mu_2},
 \mu_+=\mu_1+\mu_2$.
 Integration over $D\kappa(C)$ in (A.3) can be reduced to the
 minimization with respect to the parameter $\beta$, defining $
\mbox{\boldmath
 $ {\rm x}$}_0(\beta) =
\mbox{\boldmath
 ${\rm z}$} \beta +
\mbox{\boldmath
 ${{\rm\bar z}}$}(1-\beta), 0\leq \beta \leq
 1. $

 At large $r$ one can use the  asymptotic form of $M$, given in
 (18), then  neglecting for the moment the term $\delta$
 which is constant at large $r$ one has
 \be
 U^2\equiv \frac{M^2(z-x^0)}{2\mu_1} +
  \frac{M^2(\bar z-x^0)}{2\mu_2}=
  \sigma^2 r^2(\frac{(1-\beta)^2}{2\mu_1}+
  \frac{\beta^2}{2\mu_2})
  \ee

  Finding the stationary point of $U^2$ in $\beta$ at
  $\beta_0=\frac{\tilde \mu}{\mu_1},$ one finally obtains the
  minimized $U^2$, denoted as $<U^2>$
  \be
  <U^2>=\frac{\sigma^2r^2}{2\mu_+}.
  \ee
  Integrating  over $DR_\mu$ with boundary conditions
  $R_4(T)=T$ and $R_4(0)=0$, one has for $A$
  \be
  A=\int^T_0 dt\{\frac{m^2_1}{2\mu_1}
  +\frac{m^2_2}{2\mu_2}
  +\frac{\mu_+}{2}
  +\frac{\tilde \mu}{2}\dot{r^2}+\frac{\sigma^2 r^2}{2\mu_+}\}.
  \ee
  The corresponding Hamiltonian is
  \be
  H=\frac{m^2_1}{2\mu_1}
  +\frac{m^2_2}{2\mu_2}
  +\frac{\mu_+}{2}
  +\frac{\sigma^2 r^2}{2\mu_+}+ \frac{
\mbox{\boldmath
  ${\rm p}$}^2}{2\tilde \mu}.
  \ee
   One can vary (A.9) over $\mu_1, \mu_2$ with fixed $\tilde \mu$ to
   obtain for equal masses $m_1=m_2=m$
   \be
   H=\frac{
\mbox{\boldmath
   ${\rm  p}$}^2+m^2}{2\tilde \mu} +2\tilde \mu+\frac{\sigma^2
   r^2}{2\tilde \mu}
   \ee
   and finally varying over $\tilde \mu$ one has
   \be
   H=2\sqrt{
\mbox{\boldmath
   ${\rm p}$}^2+m^2+{\sigma^2 r^2}}
   \ee
   which yields the mass of the $q\bar q$ system
   \be
   M^2 = 8\sigma(2n_r+L+\frac32)+4m^2.
   \ee
   It is remakable that the spectrum (A.12) is the same (modulo
   factor of 4) as the WKB spectrum of Dirac equation with the linear
   confining potential [13].

   However the Regge slope of the spectrum (A.12) is
   $(8\sigma)^{-1},$ which signifies that the rotation of the string
   connecting $q$ and $\bar q$ is not properly taken into account.
   Here and above we confine ourselves to the zero angular momentum,
   leaving the problem of rotating string to subsequent publication.

        \end{document}